\documentclass[preprintnumbers,nofootinbib,showkeys,showpacs,amsmath,amssymb]{revtex4}
\usepackage{amsmath,amssymb,graphics,epsfig,subfigure}
\usepackage{color}
\usepackage{multirow}
\usepackage{booktabs}
\usepackage{changes}
\usepackage{footnote}

\usepackage{hyperref}
\hypersetup{colorlinks=true,linkcolor=blue,citecolor=magenta}

\begin{document}
	\renewcommand{\baselinestretch}{1.15}
	
	\title{Generalized Free Energy Landscapes from Iyer-Wald Formalism}
	
	\preprint{}

	\author{Shan-Ping Wu, Yu-Xiao Liu, Shao-Wen Wei\footnote{Corresponding author. E-mail: weishw@lzu.edu.cn}}

	\affiliation{$^{1}$Key Laboratory of Quantum Theory and Applications of MoE, Lanzhou Center for Theoretical Physics,
	Key Laboratory of Theoretical Physics of Gansu Province,
	Gansu Provincial Research Center for Basic Disciplines of Quantum Physics, Lanzhou University, Lanzhou 730000, China\\	
	$^{2}$Institute of Theoretical Physics $\&$ Research Center of Gravitation,
	School of Physical Science and Technology, Lanzhou University, Lanzhou 730000, China}

\begin{abstract}
The generalized free energy landscape plays a pivotal role in understanding black hole thermodynamics and phase transitions. In general relativity, one can directly derive the generalized free energy from the contributions of black holes exhibiting conical singularities. In this work, we extend this idea to general covariant theories. By employing Noether's second theorem, we present an alternative formulation of the Lagrangian, which can elucidate the role of conical singularities. We demonstrate that, in general, the contribution from conical singularities depends on the specific implementation of the regularization scheme and is not uniquely determined; this feature is explicitly exhibited and confirmed in three-dimensional new massive gravity. Nevertheless, these ambiguities can be absorbed into the second-order (and higher) corrections induced by conical singularities when the gravitational theory is described by the Lagrangian $L(g_{ab},R_{abcd})$. Moreover, for certain theories such as general relativity and Bumblebee gravity, this contribution simplifies to a well-defined result. However, the interpretation of the generalized free energy in Bumblebee gravity is somewhat different, with its extrema corresponding to the geometry of conical singularities. Our results uncover the particular properties of the generalized free energy beyond general relativity.
\end{abstract}

	\keywords{black hole, generalized free energy, covariant phase space}
	\pacs{04.70.-s, 04.70.Dy, 04.20.Fy}
	
	\maketitle
	\noindent\rule{180 mm}{0.4pt}
	\tableofcontents
	\noindent\rule{180 mm}{0.4pt}
	
	\newpage
	\section{Introduction}\label{Sec_Introduction}
	
	Semiclassical quantum effects in black hole spacetimes give rise to Hawking radiation~\cite{Hawking1975Particle}, characterized by a thermal spectrum akin to that of a blackbody, thereby revealing thermodynamic signatures intrinsic to quantum gravitational systems. Using the gravitational path integral formalism~\cite{Hartle:1976PathIntegral,Gibbons1976Action}, the dominant contribution from on-shell Euclidean black hole geometries governs the gravitational partition function, thus establishing a fundamental framework for black hole thermodynamics. Notably, the entropy of a black hole, which is fundamentally distinct from the conventional thermodynamic entropy, is proportional to the area of its event horizon~\cite{Bekenstein:1973entropy,Bardeen:1973TheFourlaws}. This seminal relationship spurred the development of the holographic principle~\cite{Susskind:1994hologram,Maldacena:1997LargeN}. On the other hand, analysis of the free energy indicates that black holes undergo phase transitions analogous to those observed in classical thermodynamic systems. The Hawking-Page phase transition~\cite{Hawking:1982AdSBH}, which characterizes the transition between Anti-de Sitter (AdS) black holes and pure AdS radiation, exhibits a remarkable correspondence to the confinement and deconfinement transitions in dual gauge theories~\cite{Witten:1998confinement,Sundborg:1999confinement}. For the charged AdS black holes, the first-order phase transitions occur between the thermodynamically small and large black hole phases, displaying critical behavior analogous to that of a van der Waals fluid~\cite{Chamblin:1999Charged,Chamblin:1999Holography,Kubiznak:2012PV}. These thermodynamic analogies provide significant insights into the microscopic degrees of freedom underlying black holes~\cite{Ruppeiner:1995,Wei:2015Microscopic,Wei:2019Repulsive,Dehyadegari:2020Microstructure,Liu:2024sfd}.
	
	Treating black holes as thermodynamic systems undergoing phase transitions raises profound questions regarding the underlying principles and kinetics of these processes. To address these issues, the concept of generalized free energy landscape has been incorporated into black hole thermodynamics, with the generalized free energy serving as a crucial tool for exploring the mechanisms underlying phase transitions~\cite{Li:2020ThermodynamicsHawkingPage,Li:2020RNAdS,Li:2021Microstructure}. This formalism extends beyond traditional on-shell descriptions by incorporating contributions from off-shell Euclidean black hole geometries. In this framework, the generalized (off-shell) free energy is constructed from the on-shell free energy $F(r_+)= M(r_+) - T_H(r_+) S(r_+)$ by replacing the Hawking temperature $T_H(r_+)$ with the ensemble temperature $T$ (where $r_+$ denotes the event horizon radius). For example, in the case of Schwarzschild AdS black holes with AdS radius $L$, one obtains the following generalized free energy
	\begin{equation}
		F_\text{off-shell}=\frac{1}{2L^2}r_+\left( r_{+}^{2}+L^2 \right) -T\pi r_{+}^{2}.
	\end{equation}	
Note that different black hole radii yield distinct free energy values, and the ensemble temperature $T$ does not necessarily equal the Hawking temperature $T_H(r_+)$. When the horizon radius $r_+$ satisfies $T_H(r_+) = T$, the generalized free energy attains a stationary point with respect to the variations in $r_+$, and it reduces to its on-shell counterpart. Moreover, building on the generalized free energy landscape, extensive discussions on black hole phases and phase transitions have emerged, providing deeper insights into various aspects of black hole thermodynamics. For instance, black hole thermodynamic topology offers a framework for classifying different types of black holes~\cite{Wei:2022Defect,Yerra:2022Topology,Fang:2022Revisiting,Wei:2024Universaltopological,Wu:2024NovelTopological}. The kinetics of phase transitions has been analyzed across diverse gravitational systems~\cite{Yang:2021KineticsKerrAdS,Wei:2020GaussBonnet,Li:2021HawkingPage,Li:2022GeneralizedFreeEnergy,Li:2023chargedGB,Liu:2023dyonicAdSBH,Fairoos:2024dGB}. Additionally, the methods based on the thermodynamic ensembles have been employed to characterize black hole phase transitions~\cite{Cheng:2024ensembleaveraged,Cheng:2024KerrAdS,Ali:2024EGB}.

	Although the definition and properties of the generalized free energy appear reasonable, substituting the Hawking temperature with the ensemble temperature to define the generalized free energy may lack a rigorous first principles foundation. However, this concern seems to have been mitigated by the ideas presented in Ref.~\cite{Li:2022GeneralizedFreeEnergy}, which demonstrated that the generalized free energy can be derived from gravitational path integrals by considering Euclidean black hole geometries with conical singularities. Using this approach, the expected form of the generalized free energy has been successfully obtained in various gravitational theories, including general relativity~\cite{Li:2022GeneralizedFreeEnergy,Liu:2023dyonicAdSBH}, Gauss-Bonnet gravity~\cite{Li:2023chargedGB}, and the massive gravity model proposed by de Rham, Gabadadze, and Tolley (dRGT)~\cite{deRham:2010kj,Fairoos:2024dGB}. Nonetheless, when extending these considerations to broader gravitational theories, several challenges and subtleties remain unresolved. In this work, we explore the path integral and generalized free energy from a more general perspective. First, we assume that the underlying theory is diffeomorphism invariant, a property intrinsic to most gravitational models. This invariance is fundamental to the formulation of black hole thermodynamics~\cite{Wald1993BHentropy,Iyer1994SomeProperties,Iyer:1995Comparison} and, as the essential gauge symmetry, allows the use of Noether second theorem to derive the gravitational action for the on-shell black hole geometry~\cite{Iyer:1995Comparison}. Second, to derive the generalized free energy, we explicitly incorporate contributions from black hole spacetimes with conical singularities. To properly account for the conical singularity, we employ the regularized function $\theta_\sigma$~\cite{Solodukhin:1994ConicalSingularity,Fursaev:1995ConicalDefects,Mann:1996Conical,Solodukhin:2011Entanglement,Nishioka:2018EntanglementEntropy}, as defined in Eq.~\eqref{eq_thetasigma}. Considering the above two points, we re-examine and reformulate the contributions of Euclidean black hole geometries with conical singularities to the gravitational action. In the general case, we find that this contribution is not straightforward and may depend on the choice of the regularized function $\theta_\sigma$. Furthermore, for two specific theories, the Bumblebee gravity~\cite{Kostelecky:1988SpontaneousBreaking,Kostelecky:2003standardmodel,Bluhm:2004Spontaneous} and new massive gravity~\cite{Bergshoeff:2009MassiveGravity,Hinterbichler:2011TheoreticalAspects,Deser:1981TopologicallyMassive,Clement:2009WarpedAdS}, we examine several related issues and discuss their implications.
	
	In this paper, we investigate the generalized free energy in gravitational theories with diffeomorphism invariance through the Euclidean path integral. In Sec.~\ref{Sec_alternativeformulation}, we re-examine an alternative formulation of the action via Noether second theorem. Sec.~\ref{Sec_GeneralizedFreeEnergy} addresses the issue of conical singularities and derives the corresponding Euclidean action. In Sec.~\ref{Sec_SpecificExamples}, we focus on the cases of Bumblebee gravity and new massive gravity. Finally, Sec.~\ref{Sec_Discussion} summarizes our findings and discusses the implications of the generalized free energy. For the sake of simplicity, we denote the volume element by the boldface symbol $\boldsymbol{\epsilon }$ and the anti-symmetric constant by the non-boldface symbol $\varepsilon$ (with $\varepsilon_{1...d} =1$). Their relationship is $\boldsymbol{\epsilon } _{a_1...a_d}\equiv\sqrt{\left| g \right|}\varepsilon _{a_1...a_d}$.
	
	\section{An Alternative Formulation of Lagrangian}\label{Sec_alternativeformulation}
	
	Gravitational actions are typically inherently characterized by diffeomorphism invariance, which is a special type of gauge symmetry. Utilizing the covariant phase space formalism, this symmetry not only naturally provides the definitions of key gravitational quantities, such as energy, angular momentum, and entropy, but also plays a crucial role in the construction of black hole thermodynamics~\cite{Wald1993BHentropy,Iyer1994SomeProperties,Iyer:1995Comparison,Papadimitriou:2005ii,Hajian:2015xlp,Seraj:2016cym,Wu:2016auq,Harlow:2019yfa,Xiao:2023lap,Hajian:2023bhq,Guo:2024oey,Gao:2023luj,Gao:2001ut}. In this section, we revisit Noether's second theorem for this gauge symmetry to derive an alternative formulation of the bulk action and present a detailed analysis, with particular emphasis on the off-shell case.
	
	To simplify our discussion, we assume that the gravitational theory is constructed solely from the metric. The Euclidean action for gravitational theories in $d$-dimension is given by
	\begin{equation}
		I_\mathrm{bulk} =-\int_M{\boldsymbol{L}\left[ g \right] },
	\end{equation}
	where $\boldsymbol{L}\left[ g \right] =L\left[ g \right] \boldsymbol{\epsilon }$, with $\boldsymbol{\epsilon }$ denoting the volume element and $L[g]$ being a scalar function of $g$ that exhibits diffeomorphism invariance. Here, we focus primarily on the contributions from the bulk action and have temporarily omitted the boundary terms. When addressing the well-posed variational problem~\cite{Gibbons1976Action,Dyer:2008BoundaryTerms,Jiang:2018SurfaceTerm} or renormalization issues~\cite{deHaro:2000HolographicReconstruction,Bianchi:2001HolographicRenormalization,Skenderis:2002LectureNotes,Guo:2025ohn}, appropriate boundary terms shall be included. By varying the bulk Lagrangian density, we obtain
	\begin{equation}
		\delta \boldsymbol{L}\left[ g \right] =\boldsymbol{E}\left[ g \right] ^{ab}\delta g_{ab}+\mathrm{d}\boldsymbol{\Theta }\left[ g,\delta g \right],
		\label{eq_deltaL}
	\end{equation}
	where $
	\boldsymbol{E}\left[ g \right] ^{ab}=E\left[ g \right] ^{ab}\boldsymbol{\epsilon }
	$ and the Euler-Lagrange equation is given by $E\left[ g \right] ^{ab} = 0$. In addition, $\mathrm{d}\boldsymbol{\Theta }\left[ g,\delta g \right]$ denotes the total derivative term arising from the variation, and $\boldsymbol{\Theta }\left[ g,\delta g \right]$ corresponds to a $(d-1)$-form that represents the presymplectic potential. When considering the infinitesimal diffeomorphism transformation generated by a vector field $\xi$, the variation of the metric is given by $\delta _{\xi} g_{ab} = \mathcal{L}_\xi g_{ab} = 2 \nabla _{(a} \xi _{b)}$. Substituting this into Eq.~\eqref{eq_deltaL}, we obtain
	\begin{equation}
		\delta _{\xi}\boldsymbol{L}\left[ g \right] =2\boldsymbol{E}\left[ g \right] ^{ab}\nabla _a\xi _b+\mathrm{d}\boldsymbol{\Theta }\left[ g,\delta _{\xi}g \right].
		\label{eq_deltaxiL}
	\end{equation}
	Taking into account the diffeomorphism invariance of the action, the above expression must be a total derivative. This requirement immediately implies the identity
	\begin{equation}
		\nabla _aE\left[ g \right] ^{ab}=0.
		\label{eq_DaEab}
	\end{equation}
	Consequently, the variation can be recast in the exact form
	\begin{equation}
		\delta _{\xi}\boldsymbol{L}\left[ g \right] =\mathrm{d}\left( \boldsymbol{M}\left[ E\left[ g \right] ,\xi \right] +\boldsymbol{\Theta }\left[ g,\delta _{\xi}g \right] \right)
		\label{eq_deltaxiL1}
	\end{equation}
	with
	\begin{equation}
		\boldsymbol{M}\left[ E\left[ g \right] ,\xi \right] _{c_1...c_{d-1}}=2E\left[ g \right] ^{ab}\xi _a\boldsymbol{\epsilon }_{bc_1...c_{d-1}}.
	\end{equation}
	This result originates from the gauge symmetry of the gravitational theory and corresponds to Noether's second theorem~\cite{Avery:2015NoetherSecondTheorem,Compere:2019AdvancedLectures}. On the other hand, the diffeomorphism invariance of the action implies that
	\begin{equation}
		\delta _{\xi}\boldsymbol{L}\left[ g \right] =\mathrm{d}\left( \xi \cdot \boldsymbol{L}\left[ g \right] \right).
		\label{eq_deltaxiL2}
	\end{equation}
	By comparing the two expressions for $\delta _{\xi}\boldsymbol{L}\left[ g \right]$, Eqs.~\eqref{eq_deltaxiL1} and \eqref{eq_deltaxiL2}, and applying the Poincar\'{e} lemma, one deduces that, locally, there exists a $(d-2)$-form field $\boldsymbol{Q}\left[g, \xi \right]$ such that
	\begin{equation}
		\boldsymbol{M}\left[ E\left[ g \right] ,\xi \right] +\boldsymbol{\Theta }\left[ g,\delta _{\xi}g \right] -\xi \cdot \boldsymbol{L}\left[ g \right] =\mathrm{d}\boldsymbol{Q}\left[g, \xi \right].
	\end{equation}
	Notably, this identity, derived without invoking to Euler-Lagrange equation $E\left[ g \right] ^{ab} = 0$, remains valid even for off-shell metrics. Thus, the bulk Lagrangian density admits an alternative formulation,
	\begin{equation}
		\sqrt{g}L\left[ g \right] \left( \xi \cdot \boldsymbol{\varepsilon } \right) =\boldsymbol{M}\left[ E\left[ g \right] ,\xi \right] +\mathbf{\Theta }\left[ g,\delta _{\xi}g \right] -\mathrm{d}\boldsymbol{Q}\left[g, \xi \right].
	\end{equation}
	Since both sides of the above equation are $(d-1)$-form, the Lagrangian density can be obtained by comparing their coefficients. In particular, if we choose $\xi = \partial_{x}$ (where $x$ is a coordinate of spacetime) and decompose the spacetime as $\mathbb{R} \times \Sigma _x $ or $\mathbb{S}^1 \times \Sigma _x $, the bulk action can be expressed as
	\begin{equation}
		I_\mathrm{bulk} [g]=\int{dx\int_{\Sigma _x}{\left( \mathrm{d}\boldsymbol{Q}\left[g, \partial_x \right] -\boldsymbol{M}\left[ E\left[ g \right] ,\partial_x \right] -\mathbf{\Theta }\left[ g,\delta _{\partial_x}g \right] \right)}}.
		\label{eq_Iexp2}
	\end{equation}
	Compared with the original Lagrangian density, this formulation is more explicit: the first term will give the boundary contribution, the second term is linear in $E[g]^{ab}$, and the third term depends on the diffeomorphism transformation and the presymplectic potential. Moreover, the second term vanishes when the metric $g$ satisfies the Euler-Lagrange equation, while the third term drops out when $\partial_x$ corresponds to an exact symmetry of the geometry. Consequently, if we take the metric $g$ to be the stationary on-shell black hole metric $g_c$ and choose $\xi$  as the Killing vector field $\xi_H$ associated with the event horizon, the on-shell Euclidean action (bulk part) simplifies to
	\begin{equation}
		I_\mathrm{bulk} [g_c]=\beta_H \left( \int_{\infty}{\boldsymbol{Q}\left[g_c, \xi _H \right]}-\int_{\mathcal{H}}{\boldsymbol{Q}\left[g_c, \xi _H \right]} \right),
	\end{equation}
	where $\beta_H$ is the inverse of Hawking temperature. In the integral, $\infty$ and $\mathcal{H}$ represent two boundaries of the hypersurface $\Sigma _x$, corresponding to spatial infinity and the black hole event horizon, respectively. At spatial infinity, the additional boundary terms must be introduced to ensure a well-posed variational principle and to implement renormalization, ultimately leading to an expression for the combined energy and angular momentum~\cite{Wald1993BHentropy,Iyer1994SomeProperties,Iyer:1995Comparison}. At the horizon, the integration yields a contribution of $S_W/\beta_H$, where $S_W$ denotes the Wald entropy~\cite{Wald1993BHentropy,Iyer1994SomeProperties,Iyer:1995Comparison}. Thus, combining them, the result for the Euclidean action is typically given by $ \beta_H (E - \Omega_H J) - S_W$, where $E$, $J$, and $\Omega_H$ denote the energy, angular momentum, and angular velocity, respectively. In this scenario, the bulk Lagrangian density for the on-shell metric $g_c$ can be written as a total derivative, so that the action receives contributions solely from the boundary values of the fields and their derivatives. In contrast, for off-shell metrics or metrics lacking exact symmetries, as indicated by Eq.~\eqref{eq_Iexp2}, the bulk action is not solely determined by the boundary integrals.
	
	In any case, the alternative formulation presented in Eq.~\eqref{eq_Iexp2} is valuable as it explicitly characterizes the contributions of the fields to the action, including those stemming from their deviations from the equations of motion and the absence of exact symmetries. In the following discussions, we utilize this formulation to analyze the contributions to the action from off-shell black holes that feature conical singularities at the horizon.
		
\section{Generalized Free Energy}\label{Sec_GeneralizedFreeEnergy}

	 The Euclidean action approach offers a successful and effective framework for constructing black hole thermodynamics~\cite{Hartle:1976PathIntegral,Gibbons1976Action}. Within this framework, the partition function is defined as
	\begin{equation}
		Z=\int{\mathcal{D} [g] \ e^{-I_E\left[ g \right]}},
		\label{eq_Zg}
	\end{equation}
	where the functional integration spans all possible metric configurations. However, an exact evaluation of this path integral is generally intractable. Among the myriad of configurations, the dominant contribution typically arises from the on-shell metric $g_c$, allowing the partition function to be approximated as
	\begin{equation}
		Z \simeq e^{-I_E\left[ g_c \right]}.
		\label{eq_Zonshell}
	\end{equation}
	This approximation not only simplifies the computation of the gravitational partition function, but also enables the derivation of the associated thermodynamic quantities. For the Euclidean black hole geometry, the ensemble temperature will impose a constraint on the geometry, requiring that the black hole metric is free of conical singularities. For example, let us consider the Schwarzschild black hole in Einstein gravity. If the ensemble temperature is $T$, the scalar curvature near the event horizon acquires an extra contribution in the form of a Dirac delta function, namely, $R\simeq \left( 8\pi MT-1 \right) \delta( r-r_+ ) $. The Euclidean geometry satisfies the on-shell condition only when the mass $M$ is fixed to $1/(8\pi T)$, thereby eliminating the conical singularity.
	
	Although the on-shell configuration plays a pivotal role in the path integral, our aim is to incorporate a broader class of geometries to capture potential quantum gravitational effects. In this study, at a fixed ensemble temperature, we analyze a class of black hole geometries characterized by conical singularities at the horizon and explore their contributions to the gravitational action. These Euclidean geometries are required to satisfy the gravitational field equations everywhere except at the horizon, where the conical singularities may appear. The free energy derived from these configurations is referred to as the generalized free energy~\cite{Li:2022GeneralizedFreeEnergy}. As discussed in Sec.~\ref{Sec_Introduction}, in some gravitational theories~\cite{Li:2022GeneralizedFreeEnergy,Liu:2023dyonicAdSBH,Li:2023chargedGB,Fairoos:2024dGB}, this generalized free energy coincides with the free energy constructed from the on-shell expression $F(r_+)= M(r_+) - T_H(r_+) S(r_+)$ after replacing the Hawking temperature $T_H(r_+)$ with the ensemble temperature $T$~\cite{Li:2020ThermodynamicsHawkingPage}. We now extend this investigation to a more general setting, namely, diffeomorphism invariant theories.

\begin{figure}[h]
		\begin{center}
			{\includegraphics[width=6cm]{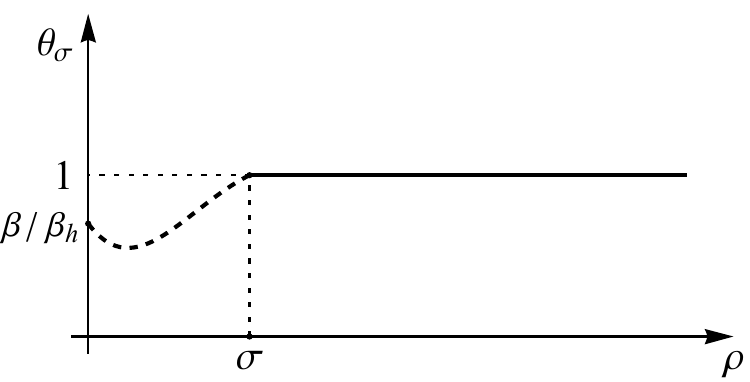}}
		\end{center}
		\caption{The functional relationship given by $\rho$ and $\theta_\sigma(\rho)$. Here the dashed bold line represents the possible curves connecting $\rho = 0$ to $\rho = \sigma$. }
		\label{FIG_ThetaFunction}
	\end{figure}

	For the geometries under consideration, the conical singularity is a critical feature that requires careful treatment. We address this issue by adopting a regularization procedure~\cite{Solodukhin:1994ConicalSingularity,Fursaev:1995ConicalDefects,Mann:1996Conical,Solodukhin:2011Entanglement,Nishioka:2018EntanglementEntropy}. The metric of a stationary black hole near the event horizon can be written as
	\begin{equation}
		ds^2= \kappa ^2\rho ^2d\tau ^2+d\rho ^2  +...,
		\label{eq_ds21}
	\end{equation}
	where $\kappa$ denotes the surface gravity. For a generic periodicity $\beta$ of Euclidean time $\tau$, the geometry near the event horizon exhibits conical singularities unless $\beta$ satisfies $\beta=\beta_H \equiv 1/T_H \equiv 2\pi/\kappa$. To address the singularity, we regularize the cone by deforming the metric so that the geometry near the event horizon (that is, at $\rho = 0$) becomes smooth while preserving the same geometry away from the horizon. To this end, we introduce a smooth function $\theta_\sigma$ that satisfies
	\begin{gather}
		\theta _{\sigma}\left( \rho \right) =1, \rho \ge \sigma ; \ \ \theta _{\sigma}\left( 0 \right) =\beta /\beta _H,
		\label{eq_thetasigma}
	\end{gather}
	with $\sigma$ a small positive real number, as illustrated in Fig.~\ref{FIG_ThetaFunction}.
	The regularized metric $g_\sigma$ is then given by
	\begin{equation}
		ds_{\sigma}^{2}= \kappa ^2\rho ^2d\tau ^2+\theta _{\sigma}\left( \rho 	\right) ^2d\rho ^2  +... \ .
		\label{eq_ds22}
	\end{equation}
	In the limit $\sigma \rightarrow 0$, the metrics in Eqs. \eqref{eq_ds21} and \eqref{eq_ds22} become nearly identical. Although the gravitational field configuration $g_\sigma$ preserves the time translation symmetry, it does not satisfy the Euler-Lagrange equations. Thus, by taking $\xi = \partial_\tau$ and using Eq.~\eqref{eq_Iexp2}, we obtain
	\begin{equation}
		I_{\mathrm{bulk}}[g_{\sigma}]=\beta \int_{\Sigma}{\left( \mathrm{d}\boldsymbol{Q}\left[ g_{\sigma},\partial _{\tau} \right] -\boldsymbol{M}\left[ E\left[ g_{\sigma} \right] ,\partial _{\tau} \right] \right)}.
	\end{equation}
	For the geometric metric configuration $g$, the Euclidean action is obtained through a limiting process, yielding
	\begin{equation}
		I_{\mathrm{bulk}}[g]=\beta \lim_{\sigma \rightarrow 0} \left( \int_{\infty}{\boldsymbol{Q}\left[ g,\partial _{\tau} \right]}-\int_{\mathcal{H}}{\boldsymbol{Q}\left[ g_{\sigma},\partial _{\tau} \right]}-\int_{\Sigma _{0\rightarrow \sigma ^+}}{\boldsymbol{M}\left[ E\left[ g_{\sigma} \right] ,\partial _{\tau} \right]} \right) .
		\label{eq_Ibulk}
	\end{equation}
	It is evident that, compared to the on-shell result, the second and third terms differ and require further treatment. Let us first consider the second term. In general, $\boldsymbol{Q}[ g_{\sigma},\partial _{\tau}]$ may include contributions involving $\theta_\sigma(0)$, $\theta_\sigma'(0)$, $\theta_\sigma''(0)$, and so on. For the contribution arising solely from $\theta_\sigma(0)$, we represent it as $S_0 (\theta_{0}(0))$. When $\theta_\sigma (0) = 1$, it should correspond to the Wald entropy; that is, $S_0(\theta _0(0)=1)=S_W$. Thus, the second term can be decomposed as
	\begin{equation}
		-\beta \lim_{\sigma \rightarrow 0} \int_{\mathcal{H}}{\boldsymbol{Q}\left[ g_{\sigma},\partial _{\tau} \right]}=-\frac{\beta}{\beta_H}S_0 (\theta_{0}(0))-\beta \Delta_S \left[ g,\theta _0(0),\theta _{0}^{\left( n \right)}(0)\left( n\ge 1 \right) \right] ,
		\label{eq_secondterm}
	\end{equation}
	where $\theta_\sigma^{(n)} (\rho)$ denotes the $n$-th derivative of $\theta_\sigma(\rho)$. For the third term in Eq.~\eqref{eq_Ibulk}, the integration may depend on $\theta_\sigma (\rho)$ and its derivatives. Specifically, one may expand $M$ as
	\begin{equation}
		\boldsymbol{M}\left[ E\left[ g_{\sigma} \right] ,\partial _{\tau} \right] =\boldsymbol{M}^{\left( 0 \right)}\left( g\left( \rho \right) ,\theta _{\sigma}\left( \rho \right) \right) +\boldsymbol{M}^{\left( 1 \right)}\left( g\left( \rho \right) ,\theta _{\sigma}\left( \rho \right) \right) \theta _{\sigma}^{\prime}\left( \rho \right) +\mathrm{other \ terms},
		\label{eq_thirdterm}
	\end{equation}
	where the contribution from the first term vanishes as $\sigma \rightarrow 0$, while the second term yields a result that depends solely on the boundary values of $\theta_{\sigma}(\rho)$. The remaining ``other terms'' are the nonlinear terms of $ \theta _{\sigma}^{\prime}\left( \rho \right)$, or functions that involve second-order or higher-order derivatives of $ \theta _{\sigma} \left( \rho \right)$. Substituting Eqs.~\eqref{eq_secondterm} and \eqref{eq_thirdterm} into Eq.~\eqref{eq_Ibulk} and adding some boundary terms, the complete Euclidean action can be expressed as
	\begin{align}
		I_{\text{off-shell}}\equiv I\left[ g \right] =&\left( \beta \int_{\infty}{\boldsymbol{Q}\left[ g,\partial _{\tau} \right]}+\text{some boundary terms} \right) -\frac{\beta}{\beta_H}S_0\left( \frac{\beta}{\beta _H} \right) -\beta \int{d\Omega \int_{\beta /\beta _H}^1{M^{\left( 1 \right)}\left( g\left( \rho =0 \right) ,\theta _{\sigma} \right) d\theta _{\sigma}}}\nonumber \\		
		&+\text{some indeterminate terms}.
		\label{eq_Ioffshell1}
	\end{align}
	The ``some boundary terms'' are constructed to ensure a well-posed variational principle~\cite{Dyer:2008BoundaryTerms,Jiang:2018SurfaceTerm} and to facilitate renormalization~\cite{deHaro:2000HolographicReconstruction,Bianchi:2001HolographicRenormalization,Skenderis:2002LectureNotes,Guo:2025ohn} (in some cases, via the background subtraction method). The ``some indeterminate terms'' correspond to $\Delta_S$ in Eq.~\eqref{eq_secondterm} and the ``other terms" in Eq.~\eqref{eq_thirdterm}, arising from the higher-order derivatives of $\theta_\sigma (\rho)$, which may be introduced by certain gravitational theories. These terms are referred to as indeterminate because the conditions (showed in Eq.~\eqref{eq_thetasigma}) imposed on the regularized function $\theta_\sigma$ are insufficient to determine the results of these terms. In other words, there are many regularized functions that satisfy condition \eqref{eq_thetasigma}, and the different regularized functions may lead to different results for ``some indeterminate terms''. It suggests that the contribution of geometries with conical singularities to the action may be ambiguous.
	
	Extracting the ``some indeterminate terms'' from Eqs. \eqref{eq_secondterm} and \eqref{eq_thirdterm}, where one arises from a volume integral and the other from a surface integral, appears somewhat inconsistent and inelegant. Instead, by partitioning the original Lagrangian density integral into two segments, we obtain a more concise result. Specifically, by decomposing the radial integration domain from 0 to $\infty$ into the intervals $[0,\sigma ^+]$ and $[\sigma ^+,\infty]$, we arrive at a concise expression
	\begin{equation}
		I_{\mathrm{bulk}}[g]=\beta \lim_{\sigma \rightarrow 0} \left( \int_{\infty}{\boldsymbol{Q}\left[ g,\partial _{\tau} \right]}-\int_{\mathcal{H} _{\sigma ^+}}{\boldsymbol{Q}\left[ g,\partial _{\tau} \right]}-\int_{\Sigma _{0\rightarrow \sigma ^+}}{\boldsymbol{L}\left[ g_{\sigma} \right]} \right).
		\label{eq_Ibulk2}
	\end{equation}
	This expression ultimately leads to the complete Euclidean action
	\begin{align}
		I_{\text{off-shell}}=&\beta \left( \int_{\infty}{\boldsymbol{Q}\left[ g,\partial _{\tau} \right]}+\text{some boundary terms} \right) -\frac{\beta}{\beta _H}S_W-\beta \lim_{\sigma^+ \rightarrow 0} \int_{\Sigma _{0\rightarrow \sigma ^+}}{\boldsymbol{L}\left[ g_{\sigma} \right]} \nonumber \\
		=&\beta \left(  \int_{\infty}{\boldsymbol{Q}\left[ g,\partial _{\tau} \right]}+\text{some boundary terms} \right) -\frac{\beta}{\beta _H}S_W-\beta \int{d\Omega \int_{\beta /\beta _H}^1{L^{\left( 1 \right)}\left( g\left( \rho =0 \right) ,\theta _{\sigma} \right) d\theta _{\sigma}}}\nonumber \\
		&+\text{some indeterminate terms}.
		\label{eq_Ioffshell2}
	\end{align}
	In this context, the term $L^{\left( 1 \right)}$ can be extracted from
	\begin{equation}
		\boldsymbol{L}\left[ E\left[ g_{\sigma} \right] ,\partial _{\tau} \right] =\boldsymbol{L}^{\left( 0 \right)}\left( g\left( \rho \right) ,\theta _{\sigma}\left( \rho \right) \right) +\boldsymbol{L}^{\left( 1 \right)}\left( g\left( \rho \right) ,\theta _{\sigma}\left( \rho \right) \right) \theta _{\sigma}^{\prime}\left( \rho \right) +\mathrm{other \ terms}.
		\label{eq_Lexp}
	\end{equation}
	Here ``some indeterminate terms'' originate from the Lagrangian density, yielding a more concise expression overall. It should be noted that the explicit parts of Eqs. \eqref{eq_Ioffshell1} and \eqref{eq_Ioffshell2} (after removing these ``some indeterminate terms'') may not coincide. The relevant calculations are provided in Appendix~\ref{Appendix_deltaI}. This apparent discrepancy does not indicate an intrinsic difference between Eqs. \eqref{eq_Ioffshell1} and \eqref{eq_Ioffshell2}; rather, it arises from different mathematical procedures, with the mismatch resulting from the presence of the indeterminate terms. In Sec.~\ref{Sec_NewMassiveGravity}, we illustrate this point with the example for new massive gravity and provide further discussion.
	
 An important supplementary remark is that when the bulk Lagrangian density is of the form $L[g] = L(g_{ab},R_{abcd})$, the relationship for the curvature and the $\theta'_\sigma(\rho)$ in Eq.~\eqref{eq_Riemmantheta} implies that the ``other terms''  in Eq.~\eqref{eq_Lexp} correspond to the contributions that are at least quadratic in $\theta'_\sigma(\rho)$. Consequently, by expanding in powers of $(1-\beta/\beta_H)$, Eq.~\eqref{eq_Ioffshell2} can be reformulated as,
	\begin{align}
		I_{\text{off-shell}}=&\beta \left( \int_{\infty}{\boldsymbol{Q}\left[ g,\partial _{\tau} \right]}+\text{some boundary terms} \right) -S_W+\left( 1-\frac{\beta}{\beta _H} \right) \left( S_W-\beta _H\int{d\Omega L^{\left( 1 \right)}\left( g\left( \rho =0 \right) ,\theta _{\sigma}=1 \right)} \right)\nonumber \\
		& +\mathcal{O} \left[ \left( 1- \beta /\beta _H \right) ^2 \right] \label{eq_Ioffshell31}
		\\
		=&\beta \left( \int_{\infty}{\boldsymbol{Q}\left[ g,\partial _{\tau} \right]}+\text{some boundary terms} \right) -S_W+\mathcal{O} \left[ \left(1- \beta /\beta _H \right) ^2 \right].
		\label{eq_Ioffshell32}
	\end{align}
The derivation of the second equality utilizes Eq.~\eqref{eq_SwL1}. A detailed derivation and explanation provided in Appendix~\ref{Appendix_deltaI}. This formula shows that when $(1 - \beta/\beta_H)$ is small (so that the $(1 - \beta/\beta_H)^k$ terms ($k>1$) can be neglected), the result is in good agreement with the construction of the off-shell energy (or off-shell Euclidean action) obtained by replacing the Hawking temperature with the ensemble temperature in the on-shell energy (or on-shell Euclidean action). However, when the ensemble temperature deviates significantly from the Hawking temperature, the result becomes less satisfactory and the actual Eq.~\eqref{eq_Ioffshell32} fails, which is mainly because that the terms beyond the first power in $(1- \beta/\beta_H)$ contribute and the hidden ambiguities become significant.
	
	In general relativity, for the charged rotating black hole with conical singularities (in Euclidean signature), the term $\boldsymbol{M}^{\left( 1 \right)}\left( g\left( \rho =0 \right) ,\theta _{\sigma} \right) $ and ``some indeterminate terms'' in Eq.~\eqref{eq_Ioffshell1} are absent. Alternatively, inspection of Eq.~\eqref{eq_Lexp} reveals that the higher-order terms in Eq.~\eqref{eq_Ioffshell32} vanish identically. In this case one obtains
	\begin{equation}
		I_\text{Einstein,off-shell}=\beta \left( M-\Omega _HJ \right) -S_W,
		\label{eq_IEinstein}
	\end{equation}
where $\beta$, $M$, $J$, and $\Omega_H$ denote the inverse temperature, energy, angular momentum, and angular velocity, respectively. This result is consistent with that result of Ref.~\cite{Li:2022GeneralizedFreeEnergy}. However, as discussed above, for general diffeomorphism invariant theories, the result is not as simple or elegant as Eq.~\eqref{eq_IEinstein}. Even for gravitational theories given by $L[g] = L(g_{ab}, R_{abcd})$, many corrections associated with conical singularity arise. Moreover, some indeterminate terms may arise, which hinder the determination of the action contribution from the black hole geometry with conical singularities. This suggests that both the off-shell action and the generalized free energy may need further refinement and investigation. To gain deeper insight, we next consider four-dimensional Bumblebee gravity and three-dimensional new massive gravity as two specific examples.
	
\section{Specific Examples}\label{Sec_SpecificExamples}

In this section, we will focus on two specific examples, Bumblebee gravity and new massive gravity.

\subsection{Bumblebee Gravity}\label{Sec_BumblebeeGravity}
	
Bumblebee gravity is a non-minimally coupled gravitational model that can describe the spontaneous Lorentz symmetry breaking~\cite{Kostelecky:1988SpontaneousBreaking,Kostelecky:2003standardmodel,Bluhm:2004Spontaneous}. Its bulk Lagrangian is given by
	\begin{equation}
		L\left[ g \right] =\frac{1}{16\pi}\left( R+\gamma B^{\mu}B^{\nu}R_{\mu \nu} 	\right) -\frac{1}{4}B_{\mu \nu}B^{\mu \nu}-V\left( B^{\mu}B_{\mu}\pm b^2 \right),
		\label{eq_LBumblebee}
	\end{equation}
	where $B_\mu$ is a massive vector field and its field strength is defined by  $B_{\mu\nu} = 2 \nabla_{[\mu} B_{\nu]}$. To trigger Lorentz symmetry breaking via a nonvanishing $B_\mu$, the potential is chosen as the form $V( B^{\mu}B_{\mu}\pm b^2 )$ with $b^2$ a positive real constant. We also impose the conditions $V(0) = V^{\prime}(0) = 0$. Thus, the non-zero vacuum expectation value of $B_\mu$ is determined by
	\begin{equation}
		B^{\mu}B_{\mu}\pm b^2 =0.
	\end{equation}
	Under a specific ansatz, an analytic solution representing a Schwarzschild-like black hole can be obtained~\cite{Casana:2017ExactSchwarzschildlike},
	\begin{equation}
		ds^2=-\left( 1-\frac{2m}{r} \right) dt^2+\frac{\ell +1}{1-\frac{2m}{r}}dr^2+r^2\left( d\theta ^2+\sin ^2\theta d\phi ^2 \right),
	\end{equation}
	with the spacelike vector field $B_a$ given by
	\begin{equation}
		B_a =b\left( 1-2m/r \right) ^{-1/2}\left( dr \right) _a.
	\end{equation}
	Here, the parameter $\ell$ is related to $\gamma$ via $\gamma =\ell /b^2$. The Hawking temperature, defined via the surface gravity, is
	\begin{equation}
		T_H= \frac{\kappa}{2\pi}=\frac{1}{4\pi r_+\sqrt{1+\ell}}.
		\label{eq_BHawking}
	\end{equation}
	Using the Iyer-Wald formalism, the energy and Wald entropy are found as
	\begin{equation}
		E=\sqrt{1+\ell} m = \frac{\sqrt{1+\ell}}{2}r_+, \ \  S_W=\pi r_{+}^{2}\left( 1+\frac{\ell}{2} \right).
		\label{eq_BWaldEnergyEntropy}
	\end{equation}
	However, the first law of thermodynamics requires that the thermodynamic temperature should be
	\begin{equation}
		T_\text{therm}=\frac{\delta E}{\delta S_W}=\frac{\sqrt{1+\ell}}{2\pi \left( 2+\ell \right) r_+},
		\label{eq_thermTemp}
	\end{equation}
	which contradicts the Hawking temperature in Eq.~\eqref{eq_BHawking}. Based on Iyer-Wald formalism, it has been shown that this discrepancy is attributed to the divergent behavior of $B_\mu$ at the event horizon~\cite{An:2024NotesOnThermodynamics}. Returning to our main topic, this inconsistency prompts an investigation into the generalized free energy landscape, wherein temperature is interpreted as the ensemble temperature. For black holes that extremize the generalized free energy, a temperature can be derived. This raises the question of whether this temperature corresponds to $T_H$ or $T_\text{therm}$, a matter of considerable interest.
	
	Next, we employ the path integral approach to derive the generalized free energy. In the context of Bumblebee gravity, we find that for the regularized metric the quantity $\boldsymbol{M}\left[ E\left[ g_{\sigma} \right] ,\partial _{\tau} \right] $ vanishes, and there is no contribution from the derivative of $\theta_\sigma (\rho)$ in $\boldsymbol{Q}\left[ g_{\sigma},\partial _{\tau} \right] $. Consequently, the contributions from $\boldsymbol{M}^{\left( 1 \right)}$ and ``some indeterminate terms'' in Eq.~\eqref{eq_Ioffshell1} vanish. This aspect is similar to general relativity. Therefore, the contribution from the horizon part in Eq.~\eqref{eq_Ioffshell1} can be written as
	\begin{align}
		-\frac{\beta}{\beta _H}S_0\left( \frac{\beta}{\beta _H} \right) -\beta \int{d\Omega \int_{\beta /\beta _H}^1{M^{\left( 1 \right)}\left( g\left( \rho =0 \right) ,\theta _{\sigma} \right) d\theta _{\sigma}}}+\text{some indeterminate terms} =-S_W.
	\end{align}
	To ensure a well-posed variational principle in this gravitational theory, one must include the boundary term
	\begin{equation}
		I_{\mathrm{surf}}=\frac{1}{8\pi}\int_{\partial M}{d^3 x \sqrt{h}\left( K+\frac{1}{2}\gamma B^aB^b\left( K_{ab}-n_an_bK \right) \right)} +		\int_{\partial M}{d^3 x\sqrt{h}n_{a}B_{b}B^{a b}}.
	\end{equation}
	Incorporating this term into the ``some boundary terms'' in Eq.~\eqref{eq_Ioffshell1} reveals that the contribution at infinity (the term in the first parentheses in Eq.~\eqref{eq_Ioffshell1}) is divergent. This divergence necessitates renormalization, for which we adopt the background subtraction method. Ultimately, the off-shell Euclidean action can be calculated as
	\begin{equation}
		I_\text{off-shell}=\beta \sqrt{\ell +1}m-S_W = \beta E- S_W,
	\end{equation}
	and the corresponding generalized (off-shell) free energy is
	\begin{equation}
		F_\text{off-shell} = \frac{I_\text{off-shell}}{\beta}=E-\frac{1}{\beta}S_W=\frac{\sqrt{1+\ell}}{2}r_+-\frac{1}{\beta}\pi r_{+}^{2}\left( 1+\frac{\ell}{2} \right).
	\end{equation}
	Moreover, the black hole energy and entropy can also be obtained from $\partial _{\beta}I_\text{off-shell}$ and $	-\left( 1-\beta \partial _{\beta} \right) I_\text{off-shell}$, respectively, yielding results consistent with Eq.~\eqref{eq_BWaldEnergyEntropy}. Taking the variation of the generalized free energy $F_g$ with respect to the event horizon radius $r_+$ gives
	\begin{equation}
		r_+=\frac{\sqrt{1+\ell}}{2\pi \left( 2+\ell \right) T}.
	\end{equation}	
	This extremal condition coincides with the thermodynamic temperature in Eq.~\eqref{eq_thermTemp}, yet it does not match the Hawking temperature in Eq.~\eqref{eq_BHawking}. This indicates that the contribution at the extremum is not evaluated by the on-shell solution but is determined by a geometry that possesses conical singularities. This situation is different from general relativity. Finally, the deficit angle of the Euclidean black hole corresponding to the free energy extremum is given by
	\begin{equation}
		\text{deficit angle} = 2\pi \left( 1-\frac{\beta}{\beta _H} \right) =\frac{\pi \ell}{1+\ell}.
	\end{equation}
	This deficit angle is related to Lorentz symmetry breaking; when $\ell$ vanishes the metric reduces to that of the Schwarzschild black hole and the deficit angle disappears.

\subsection{New Massive Gravity}\label{Sec_NewMassiveGravity}

	In this subsection, we examine the Euclidean action for geometries with the conical singularities in higher-order gravity theories. One simple example of such theories is three-dimensional new massive gravity~\cite{Bergshoeff:2009MassiveGravity,Hinterbichler:2011TheoreticalAspects,Deser:1981TopologicallyMassive,Clement:2009WarpedAdS}. The action we consider is given by
	\begin{equation}
		L\left[ g \right] =\frac{1}{16\pi}\left( R-2\Lambda -\alpha \left( \frac{3}{8}R^2-R_{\mu \nu}R^{\mu \nu} \right) \right).
	\end{equation}
	A special class of solutions, analogous to the BTZ black hole~\cite{Clement:2009WarpedAdS}, is described by the following line element
	\begin{equation}
		ds^2=-\left( -m+\frac{r^2}{\ell ^2}+\frac{j^2}{4r^2} \right) dt^2+\frac{1}{\left( -m+\frac{r^2}{\ell ^2}+\frac{j^2}{4r^2} \right)}dr^2+r^2\left( d\phi ^2-\frac{j}{2r^2}dt \right) ^2,
	\end{equation}
	where the length scale $\ell$ is determined by the relation $\Lambda =-1/\ell ^2+\alpha /\left( 4\ell ^4 \right) $. This black hole possesses two horizons with radii $r_{\pm}=\ell \sqrt{\frac{1}{2}m\pm \frac{1}{2}\sqrt{m^2-j^2/\ell ^2}}$. For the event horizon, the relevant Killing vector field is $\xi_H = \partial_t + \Omega_H \partial_\phi$ with $\Omega _H=j/(2r_{+}^{2})$. The Hawking temperature is obtained from the surface gravity,
	\begin{equation}
		T_H = \frac{\kappa}{2\pi}=\frac{4r_{+}^{4}-\ell ^2j^2}{8\pi \ell ^2r_{+}^{3}}.
	\end{equation}
	Using the Iyer-Wald formalism, the energy, angular momentum, and Wald entropy are given by
	\begin{equation}
		E=\frac{1}{8}m\left( 1+\frac{\alpha}{2\ell ^2} \right), \ \ J=\frac{1}{8}j\left( 1+\frac{\alpha}{2\ell ^2} \right), \ \   S_W=\frac{\pi r_+}{2}\left( 1+\frac{\alpha}{2\ell ^2} \right),
	\end{equation}
	corresponding to the Killing vector fields $\partial_t$, $\partial_\phi$, and $\xi_H$, respectively. To ensure the well-posed variational principle, we introduce the boundary term
	\begin{equation}
		I_{\mathrm{surf}}=\frac{1}{8\pi}\int_{\partial M}{d^3x\sqrt{h}\left( K-\beta \left( \frac{3}{4}K-R^{ab}\left( K_{ab}-n_an_bK \right) \right) \right)}.
		\label{eq_NMGSurf}
	\end{equation}
	The ``some boundary terms'' in Eq. \eqref{eq_Ioffshell1} or \eqref{eq_Ioffshell2} include the contributions from Eq.~\eqref{eq_NMGSurf}, as well as extra terms arising from the background subtraction method. With these in place, one finds
	\begin{equation}
		\left( \beta \int_{\infty}{\boldsymbol{Q}\left[ g,\partial _{\tau} \right]}+\text{some boundary terms} \right) =\beta \frac{\left( 4r_{h}^{4}-j^2\ell ^2 \right) \left( \alpha +2\ell ^2 \right)}{64r_{h}^{2}\ell ^4}=\beta \left( E-\Omega _H J \right).
	\end{equation}
	This result is obtained at spatial infinity. At the horizon, one can find that $\boldsymbol{Q}\left[ g_{\sigma},\partial _{\tau} \right] $ contains the function $\theta_\sigma$ and its derivatives. From Eq.~\eqref{eq_secondterm}, we can isolate the term that only depends on the value of the function $\theta_\sigma$, which we denote as the function
	\begin{equation}
			S_0\left( \theta \right) =\frac{1}{2}\pi r_+\left( \frac{1}{\theta}+\frac{\alpha}{2\ell ^2\theta ^3} \right).
	\end{equation}
	Obviously, setting $\theta=1$ recovers $S_0 (1) = S_W$. For the complete action of the black hole with conical singularities, we obtain
	\begin{align}
			I_\text{off-shell} =&\beta \left( M-\Omega _H J \right) -\frac{1}{2}\pi r_+\left( 1+\frac{\alpha}{2\ell ^2} \right) +\frac{\alpha}{12\ell ^2}\left( 3r_+\left( 1-\frac{\beta _{H}^{2}}{\beta ^2} \right) +2\pi \frac{\ell ^2}{\beta _H}\left( \frac{\beta}{\beta _H}-\frac{\beta _{H}^{2}}{\beta ^2} \right) \right)\nonumber \\
			&+\text{some indeterminate terms},
			\label{eq_Ioffshell1NMG}
	\end{align}
from Eq.~\eqref{eq_Ioffshell1},	whereas
	\begin{align}
		I_{\text{off-shell}} =&	\beta \left( M-\Omega _HJ \right) -\frac{1}{2}\pi r_+\left( 1+\frac{\alpha}{2\ell ^2} \right) -\frac{\alpha \pi}{12\ell ^2}r_+\left( 1+2\frac{\beta}{\beta _H} \right) \left( 1-\frac{\beta _H}{\beta} \right) ^2\nonumber	 \\
		&+\text{some indeterminate terms},
		\label{eq_Ioffshell2NMG}
	\end{align}
	from Eq.~\eqref{eq_Ioffshell2}.	Excluding the ``some indeterminate terms'', the discrepancy between these two expressions arises because the contributions from higher-order derivatives of $\theta_\sigma$ are sensitive to the chosen regularization scheme. As discussed in Appendix~\ref{Appendix_deltaI}, these two distinct results reflect different ways of handling the indeterminacy. Moreover, the presence of ``some indeterminate terms'' suggests that, in three-dimensional new massive gravity, contributions from conical singularities cannot be entirely regularized and uniquely determined. From another perspective, to clearly determine the contributions from conical singularities, defining the regularized function $\theta_\sigma$ solely by its values at $\rho = 0$ and $\rho =\sigma$ (as given in Eq.~\eqref{eq_thetasigma}) is insufficient; additional derivative values, or a specific form for the regularized function, should be provided. However, imposing such conditions appears unphysical, warranting further investigation. In addition, given that the action for new massive gravity takes the form $L(g_{ab}, R_{abcd})$, the indeterminate terms in Eq.~\eqref{eq_Ioffshell2NMG} are at least quadratic in $(1-\beta/\beta_H)$ (some detailed analysis can be found in the Appendix~\ref{Appendix_deltaI}). That is, Eq.~\eqref{eq_Ioffshell2NMG} can be expressed as
	\begin{equation}
		I_{\text{off-shell}} =	\beta \left( M-\Omega _HJ \right) -\frac{1}{2}\pi r_+\left( 1+\frac{\alpha}{2\ell ^2} \right) +\mathcal{O} \left[ \left( 1-\beta /\beta _H \right) ^2 \right],
	\end{equation}
	which is consistent with Eq.~\eqref{eq_Ioffshell32}. Correspondingly, the generalized free energy should be
	\begin{equation}
		F_{\text{off-shell}} =	 \left( M-\Omega _HJ \right) - \frac{1}{\beta}\frac{1}{2}\pi r_+\left( 1+\frac{\alpha}{2\ell ^2} \right) +\mathcal{O} \left[ \left( 1-\beta /\beta _H \right) ^2 \right].
	\end{equation}
	It can be seen that, since the higher-order terms in the above expression are expanded in powers of $(1-\beta/\beta_H)^2$ and higher, the black hole radius $r_+$ satisfying $\beta_H = \beta$ remains an exact stationary point of the generalized free energy.
	
\section{Discussion and Conclusion}\label{Sec_Discussion}

In black hole thermodynamics, the generalized (or off-shell) free energy of black holes is a topic of significant interest. Nevertheless, the generalized free energy of black holes has lacked a more profound foundational basis until the seminal work in Ref.~\cite{Li:2022GeneralizedFreeEnergy}, which incorporated the contribution of Euclidean black holes exhibiting conical singularities. Basing on this idea, our work extends the analysis to encompass more general covariant theories.

We addressed this issue from the gauge symmetry perspective and emphasized an alternative formulation of the bulk action given in Eq.~\eqref{eq_Iexp2}. In order to address the conical singularities on the Euclidean black hole horizon, we employ a regularization method based on a function $\theta_{\sigma}$ that satisfies Eq.~\eqref{eq_thetasigma}. Our study not only reveals the impact of conical singularities on the Euclidean gravitational action, but also demonstrates that the definition \eqref{eq_thetasigma} of the regularized function is insufficient to fully capture these contributions; indeed, the final outcome depends on the specific construction of the regularized function. This observation suggests that further investigation is needed and implies that black holes with conical singularities may not naturally provide a robust foundation for defining the generalized free energy. Nevertheless, although this ambiguity may arise in general theories, it can be avoided in certain specific cases. For example, in general relativity, the regularized function $\theta_{\sigma}$ is adequate, and the generalized free energy can be explicitly and uniquely determined. Furthermore, for theories defined by the action $L(g_{ab},R_{abcd})$, when considering the power series expansion in $(1-\beta/\beta_H)$, the undetermined terms can be absorbed into contributions of second-order and higher.
	
	Extending our analysis, we examined the case of Bumblebee gravity. In this theory the contribution of conical singularities to the action is uniquely determined, and the Wald entropy plays a crucial role in constructing the generalized free energy. Unlike general relativity, the presence of additional fields $B_\mu$ in Bumblebee gravity introduces a discrepancy between the thermodynamic temperature and the Hawking temperature. Under a fixed ensemble temperature, extremizing the generalized free energy yields a black hole temperature that exactly coincides with the thermodynamic temperature. This suggests that the extremum of the generalized free energy is determined by the Euclidean black hole geometry with conical singularities, highlighting the thermodynamic significance of such geometries.
	
	Additionally, we investigated the three-dimensional new massive gravity as a representative example of the higher-order gravitational theory. Our results confirm that the regularization via the regularized function may not fully determine the contribution of conical singularities to the action. In higher-order theories, higher derivatives of the regularized function $\theta_\sigma$ tend to appear in the expressions $\boldsymbol{Q}\left[ g_{\sigma},\partial _{\tau} \right] $, $
	\boldsymbol{M}\left[ E\left[ g_{\sigma} \right] ,\partial _{\tau} \right]$, and $\boldsymbol{L}\left[ g_{\sigma} \right] $. Consequently, in most cases, the contribution of conical singularities to the action cannot be fully determined by the definition in Eq.~\eqref{eq_thetasigma}.  A similar discussion can also be found in Ref.~\cite{Fursaev:1995ConicalDefects}.
	
	Returning to the original definition~\cite{Li:2020ThermodynamicsHawkingPage,Li:2020RNAdS,Li:2021Microstructure}, it appears natural to obtain the generalized free energy by replacing the Hawking temperature in the on-shell free energy with the ensemble temperature. Within this framework the minimum (maximum) of the generalized free energy corresponds to the thermodynamic state with positive (negative) heat capacity. Furthermore, in several specific gravitational theories, including Einstein gravity~\cite{Li:2022GeneralizedFreeEnergy,Liu:2023dyonicAdSBH}, Gauss-Bonnet gravity~\cite{Li:2023chargedGB}, and dRGT massive gravity~\cite{Fairoos:2024dGB}, the generalized free energy obtained from black hole geometries with conical singularities is consistent with this original definition. However, in the two gravitational theories considered here, Bumblebee gravity and new massive gravity, the results deviate from the original definition or interpretation. For theories constructed by the Lagrange form $L(g_{ab}, R_{abcd})$ (including new massive gravity), the generalized free energy can retain its original definition only when these terms of order $(1-\beta/\beta_H)^2$ and higher are neglected. In any case, these results imply that the connection between the Euclidean path integral and the generalized free energy warrants further scrutiny. Additionally, it is also worth noting that Appendix C of Ref.~\cite{Liu:2023dyonicAdSBH} provides an alternative method to handle the Euclidean black hole geometry with conical singularities. Instead of regularizing the singularities, this approach introduces a boundary near the event horizon to exclude the conical singularities and imposes additional boundary terms. Under this consideration, the expected off-shell action and the generalized free energy are obtained. Nevertheless, the additional boundary term appears to be theory dependent and may also require further investigation.
	
	Finally, we emphasized that the alternative formulation provided in Eq.~\eqref{eq_Iexp2} is noteworthy because it explicitly captures the contribution of the off-shell metric to the gravitational action. This formulation facilitates a clearer analysis of contributions to the action arising from deviations in the equations of motion or from departures from certain symmetries. Although the resulting expression is more complex than the conventional Lagrangian density, it may serve as a useful representation for studying the quantum effects of gravity.

\section*{Acknowledgments}

This work was supported by the National Natural Science Foundation of China (Grants No. 12475055, No. 12475056, No. 12247101), the National Key
Research and Development Program of China (Grant No. 2021YFC2203003), and the 111 Project under (Grant No. B20063), the Gansu Province’s Top Leading Talent Support Plane.

	\appendix
	\section{Supplementary Calculations}\label{Appendix_deltaI}
	In this Appendix, we provide supplementary calculations and additional discussions that augment the analysis presented in the main text.
	In Sec.~\ref{Sec_GeneralizedFreeEnergy}, we showed that Eqs. \eqref{eq_Ioffshell1} and \eqref{eq_Ioffshell2} are not equal after removing ``some indeterminate terms''. Here, we provide a detailed calculation to clarify this discrepancy. To that end, consider the first and third terms of Eq.~\eqref{eq_Ibulk2}. For the configuration $g_\sigma$, one obtains
	\begin{equation}
		-\beta \int_{\mathcal{H} _{\sigma ^+}}{\boldsymbol{Q}\left[ g,\partial _{\tau} \right]}-\beta \int_{\Sigma _{0\rightarrow \sigma ^+}}{\boldsymbol{L}\left[ g_{\sigma} \right]}=-\beta \int_{\mathcal{H} _{\sigma ^+}}{\boldsymbol{Q}\left[ g,\partial _{\tau} \right]}+\beta \int_{\Sigma _{0\rightarrow \sigma ^+}}{\mathrm{d}\boldsymbol{Q}\left[ g_{\sigma},\xi \right]}-\beta \int_{\Sigma _{0\rightarrow \sigma ^+}}{\boldsymbol{M}\left[ E\left[ g_{\sigma} \right] ,\xi \right]}.
	\end{equation}
	If the Lagrangian $\boldsymbol{L}$ on the left-hand side is taken as $\boldsymbol{L}^{(1)}$, this result corresponds to Eq.~\eqref{eq_Ioffshell2}. On the other hand, if the term $\boldsymbol{M}$ on the right hand side is replaced by $\boldsymbol{M}^{(1)}$ and the contribution $\Delta _S $ (defined in Eq.~\eqref{eq_secondterm}) is subtracted from the right-hand side, the result expression corresponds to Eq.~\eqref{eq_Ioffshell1}. Hence, apart from ``some undetermined terms'', the difference between Eqs. \eqref{eq_Ioffshell1} and \eqref{eq_Ioffshell2} is given by
	\begin{align}
		\Delta I&=\beta \left( \int_{\Sigma _{0\rightarrow \sigma ^+}}{\mathrm{d}\boldsymbol{Q}\left[ g_{\sigma},\xi \right]}+\Delta _S\left[ g,\theta _0(0),\theta _{0}^{\left( n \right)}(0)\left( n\ge 1 \right) \right] -\int_{\Sigma _{0\rightarrow \sigma ^+}}{\left( \boldsymbol{M}^{\left( 1 \right)}\left( g\left( \rho \right) ,\theta _{\sigma}\left( \rho \right) \right) -\boldsymbol{L}^{\left( 1 \right)}\left( g\left( \rho \right) ,\theta _{\sigma}\left( \rho \right) \right) \right) \theta _{\sigma}^{\prime}\left( \rho \right)} \right) \nonumber
		\\
		&=\beta \left( \int_{\Sigma _{0\rightarrow \sigma ^+}}{\mathrm{d}\boldsymbol{Q}\left[ g_{\sigma},\xi \right]}-\int_{\Sigma _{0\rightarrow \sigma ^+}}{\left( \mathrm{d}\boldsymbol{Q} \right) ^{\left( 1 \right)}\left( g\left( \rho \right) ,\theta _{\sigma}\left( \rho \right) \right) \theta _{\sigma}^{\prime}\left( \rho \right)}+\Delta _S\left[ g,\theta _0(0),\theta _{0}^{\left( n \right)}(0)\left( n\ge 1 \right) \right] \right).
		\label{eq_DeltaI}
	\end{align}
	Here, $( \mathrm{d}\boldsymbol{Q}\left[ g_{\sigma},\xi \right] ) ^{(1)} $ denotes the coefficient of $\theta _{\sigma}^{\prime}\left( \rho \right)$  in the expansion
	\begin{equation}
		\mathrm{d}\boldsymbol{Q}\left[ g_{\sigma},\xi \right] =\left( \mathrm{d}\boldsymbol{Q} \right) ^{\left( 0 \right)}\left( g\left( \rho \right) ,\theta _{\sigma}\left( \rho \right) \right) +\left( \mathrm{d}\boldsymbol{Q} \right) ^{\left( 1 \right)}\left( g\left( \rho \right) ,\theta _{\sigma}\left( \rho \right) \right) \theta _{\sigma}^{\prime}\left( \rho \right) +\mathrm{other \ terms}.
		\label{eq_dQexpand}
	\end{equation}
 	A detailed analysis shows that the difference $\Delta I$ does not necessarily vanish. If $\boldsymbol{Q}\left[ g_{\sigma},\xi \right]$ contains no terms involving derivatives of $\theta _{\sigma}(\rho)$, the $\Delta  _S$ term is absent and the first two integrals in Eq.~\eqref{eq_DeltaI} cancel each other. However, if $\boldsymbol{Q}\left[ g_{\sigma},\xi \right]$ includes a term of the form $
 	f\left( \rho ,\theta _{\sigma}(\rho ) \right) \theta _{\sigma}^{\prime}(\rho )
 	$ for some function $f$, then the cancellation between the first and third terms occurs, while the second term may provide a nonvanishing contribution. Therefore, in general, Eqs. \eqref{eq_Ioffshell1} and \eqref{eq_Ioffshell2} are not equivalent. Furthermore, a closer examination reveals that in Eq.~\eqref{eq_Ioffshell1}, the total derivative term $\mathrm{d}\boldsymbol{Q}\left[ g_{\sigma},\xi \right]$ in the Lagrangian density is smoothly matched at $\rho = \sigma$, and the entire indeterminacy of $\boldsymbol{Q}\left[ g_{\sigma},\xi \right]$ is attributed solely to the horizon ($\rho = 0$) contribution. In contrast, in Eq.~\eqref{eq_Ioffshell2}, the term $\mathrm{d}\boldsymbol{Q}\left[ g_{\sigma},\xi \right]$ accounts only for the first two terms in the expansion of Eq.~\eqref{eq_dQexpand},  thereby ascribing the indeterminacy to the contribution between $\rho = \sigma$ and the horizon ($\rho = 0$). These results reflect two distinct approaches for handling the indeterminacy.
 	
 For the derivation from Eq.~\eqref{eq_Ioffshell2} to Eq.~\eqref{eq_Ioffshell32}, we provide a detailed explanation here. First, for the metric $g$ and deformed metric $g_\sigma$, their curvature relationship near the horizon is given in Refs.~\cite{Solodukhin:1994ConicalSingularity,Fursaev:1995ConicalDefects}. For this key result, in our case, $\delta_\sigma$ should transform to $\theta^\prime_\sigma$. Furthermore, the two normal vectors $n_1$ and $n_2$ are combined into the binormal $n$ ($n = n_1\wedge n_2$) associated with the horizon, yielding the curvature relation near the horizon as follows,
 	\begin{equation}
 		R{\left[ g_{\sigma} \right] ^{ab}}_{cd} \approx R{\left[ g \right] ^{ab}}_{cd}+\frac{2\pi}{\beta _H}n^{ab}n_{cd}\theta _{\sigma}^{\prime}\left( \rho \right).
 		\label{eq_Riemmantheta}
 	\end{equation}
 	The symbol ``$\approx$'' indicates that the left and right hand expressions are equal when the value of $\theta_\sigma(\rho)$ is set to 1. If we consider the Lagrangian density in the form $L[g] = L(g_{ab}, R_{abcd})$,  then near the horizon, we have
 	\begin{equation}
 		L\left[ g_{\sigma} \right] =L\left( {g_{\sigma}}_{ab},R\left[ g_{\sigma} \right] _{abcd} \right) \approx L\left( {g}_{ab},R\left[ g \right] _{abcd} \right) +\frac{2\pi}{\beta _H}\frac{\partial L\left( {g}_{ab},R\left[ g \right] _{abcd} \right)}{\partial R_{abcd}}n_{ab}n_{cd}\theta _{\sigma}^{\prime}\left( \rho \right) +\mathcal{O} \left( \theta _{\sigma}^{\prime}\left( \rho \right) ^2 \right) .
 		\label{eq_Ltheta}
 	\end{equation}
 	Comparing with Eq.~\eqref{eq_Lexp}, it follows that the second term in the above expression corresponds to $L^{(1)}$. Thus, we obtain the following relation
 	\begin{equation}
 		\beta _H\int{d\Omega \boldsymbol{L}^{\left( 1 \right)}\left( g\left( \rho =0 \right) ,\theta _{\sigma}(\rho)=1 \right)}=\beta _H\int{d\Omega \frac{2\pi}{\beta _H}\frac{\partial L\left( g_{ab},R\left[ g \right] _{abcd} \right)}{\partial R_{abcd}}n_{ab}n_{cd}}=2\pi \int{d\Omega \frac{\partial L\left( g_{ab},R\left[ g \right] _{abcd} \right)}{\partial R_{abcd}}n_{ab}n_{cd}}.
 	\end{equation}
 	The final expression corresponds to the Wald entropy $S_W$,  leading to
 	\begin{equation}
 		 \beta _H\int{d\Omega L^{\left( 1 \right)}\left( g\left( \rho =0 \right) ,\theta _{\sigma}=1 \right)} = S_W.
 		 \label{eq_SwL1}
 	\end{equation}
 	For this part, similar discussions can also be found in Refs.~\cite{Azeyanagi:2007NearExtremal,Myers:2010ctheorems,Solodukhin:2011Entanglement}. Returning to the off-shell action expression in Eq.~\eqref{eq_Ioffshell2} and expanding it in powers of $(1-\beta/\beta_H)$, we obtain
 	\begin{align}
 		I_{\text{off-shell}}=&\beta \left(  \int_{\infty}{\boldsymbol{Q}\left[ g,\partial _{\tau} \right]}+\text{some boundary terms} \right) -\frac{\beta}{\beta _H}S_W-\beta \int{d\Omega \int_{\beta /\beta _H}^1{L^{\left( 1 \right)}\left( g\left( \rho =0 \right) ,\theta _{\sigma} \right) d\theta _{\sigma}}}\nonumber \\
 		&+\text{some indeterminate terms}\label{eq_Iline1}
 		 \\
 		= &\beta \left(  \int_{\infty}{\boldsymbol{Q}\left[ g,\partial _{\tau} \right]}+\text{some boundary terms} \right) -\frac{\beta}{\beta _H}S_W-\beta \int{d\Omega \int_{\beta /\beta _H}^1{L^{\left( 1 \right)}\left( g\left( \rho =0 \right) ,\theta _{\sigma} \right) d\theta _{\sigma}}}\nonumber \\
 		&+ \mathcal{O} \left[ \left(1- \beta /\beta _H \right) ^2 \right] \label{eq_Iline2}
 		\\
 		= &\beta \left( \int_{\infty}{\boldsymbol{Q}\left[ g,\partial _{\tau} \right]}+\text{some boundary terms} \right) -S_W+\left( 1-\frac{\beta}{\beta _H} \right) \left( S_W-\beta _H\int{d\Omega L^{\left( 1 \right)}\left( g\left( \rho =0 \right) ,\theta _{\sigma}(\rho)=1 \right)} \right)\nonumber \\
 		& +\mathcal{O} \left[ \left( 1- \beta /\beta _H \right) ^2 \right] \label{eq_Iline3}
 		\\
 		=&\beta \left( \int_{\infty}{\boldsymbol{Q}\left[ g,\partial _{\tau} \right]}+\text{some boundary terms} \right) -S_W+\mathcal{O} \left[ \left(1- \beta /\beta _H \right) ^2 \right]. \label{eq_Iline4}
 	\end{align}
 	The ``some indeterminate terms" here are obtained from the linear combination of terms higher than first power in $\theta^{\prime}_{\sigma}(\rho)$ in Eq.~\eqref{eq_Ltheta}. Noting that $\theta^{\prime}_{\sigma}(\rho) \sim (1-\beta/\beta_H)$, Eq.~\eqref{eq_Iline1} can be converted into Eq.~\eqref{eq_Iline2}. By expanding Eq.~\eqref{eq_Iline2} in powers, we obtain Eq.~\eqref{eq_Iline3}. Finally, by employing the relation \eqref{eq_SwL1}, the final result Eq.~\eqref{eq_Iline4} is derived.

	\bibliographystyle{unsrt}

\end{document}